\documentclass[aps,prd,twocolumn,epsf,nofootinbib,eqsecnum]{revtex4}
\usepackage{amsfonts,latexsym,eucal,amsmath,amssymb,times,mathrsfs}
\usepackage[dvips]{graphicx}
\usepackage{color}

\newcommand{\bm}[1]{\hbox{\boldmath{$#1$}}}
\newcommand{\sbm}[1]{\hbox{\boldmath{\scriptsize$#1$}}}
\newcommand{\bk}{\mbox{\boldmath{\scriptsize $k$}}}

\newcommand{\Mp}{M_{\rm pl}}
\newcommand{\dd}{{\rm d}}
\newcommand{\sR}{{^s\!R}}

\newcommand{\gR}{{^g\!R}}
\newcommand{\gz}{{^g\!\zeta}}
\newcommand{\Approx}{\stackrel{\tiny\rm IR}{\approx}}

\begin{document}

\thispagestyle{empty}


\title{Influence of gauge artifact on adiabatic and entropy
perturbations during inflation}
\date{\today}
\author{Yuko Urakawa$^{1,2}$}
\email{yurakawa_at_ffn.ub.es}
\address{\,\\ \,\\
$^{1}$ Departament de F{\'\i}sica Fonamental i Institut de Ci{\`e}ncies del Cosmos, 
Universitat de Barcelona,
Mart{\'\i}\ i Franqu{\`e}s 1, 08028 Barcelona, Spain\\
$^{2}$ Department of Physics, Ochanomizu University 2-1-1 Otsuka, Bunkyo, Tokyo, 112-8610 Japan}



\begin{abstract}
In recent publications we proposed one way to calculate
gauge-invariant variables in the local observable universe, which is limited to
 a portion of the whole universe. To provide a theoretical prediction of
 the observable fluctuations, we need to preserve the gauge-invariance
 in the local universe, which is referred to the genuine gauge-invariance.  
The importance of the genuine gauge-invariance is highlighted in the
 study of primordial fluctuations in the infrared (IR) limit. The
 stability against IR loop corrections to primordial fluctuations is
 guaranteed in requesting the genuine gauge-invariance. The genuine
 gauge-invariance also gives the impact on the detectable
 fluctuations. We showed that at observable scales the bi-spectrum calculated in the conventional
 perturbation theory vanishes in the squeezed
 limit, if we request the genuine gauge invariance. This indicates that the conventional
 bi-spectrum is, in this limit, dominated by a gauge artifact, which cannot be observed.
These studies have been elaborated in single-field 
 models of inflation. In this paper we generalize our argument to multi-field models
 of inflation, where, in addition to the adiabatic field, the entropy
 field can participate in the generation of primordial fluctuations. We
 will find that the entropy field can generate the observable fluctuations,
 which cannot be eliminated by gauge transformations in the local universe. 
\end{abstract}


\pacs{98.80.-k, 98.80.Bp, 98.80.Cq, 04.20.Cv}
\maketitle


\section{Introduction}
The conventional cosmological perturbation theory has been established based on
the assumption that we know the whole spatial region of the
universe with infinite volume. We should, however, recognize that this
does not meet the case in the actual observations, because the observable
portion of the universe is limited. In our recent works~\cite{IRsingle,
IRgauge_L, IRgauge, IRNG}, we pointed out
the necessity of distinguishing the gauge invariance in the whole
universe with infinite volume from that in the local universe with finite
volume. To preserve the gauge-invariance in the whole universe, it is
sufficient to request the invariance under normalizable gauge transformations, which
become regular at spatial infinity. However, in the local universe,
where we need not to concern about the regularity at infinity, it is necessary to request
the invariance under both normalizable and non-normalizable
gauge transformations. We
discriminate gauge-invariant quantities that can be constructed in our
local observable universe, referring them as {\it genuine gauge-invariant variables}.
The observable fluctuations should be such a genuinely gauge-invariant variable.

The genuine gauge-invariant perturbation has a large impact
namely on the infrared (IR) modes of fluctuations. The adiabatic vacuum, which yields the scale-invariant spectrum,
is supposed to be a natural vacuum in the inflationary
universe. However, once the interaction turns on, it is not manifest whether
the adiabatic vacuum is stable or unstable against the IR contributions in loop
corrections~\cite{Boyanovsky:2004gq, Boyanovsky:2004ph, Boyanovsky:2005sh,
Boyanovsky:2005px, Tsamis:1996qm, Tsamis:1996qq, Onemli:2002hr, Brunier:2004sb, Prokopec:2007ak,
Sloth:2006az, Sloth:2006nu, Seery:2007we, Seery:2007wf, Urakawa:2008rb,
Adshead:2008gk, Cogollo:2008bi, Rodriguez:2008hy, Seery:2009hs,
Gao:2009fx, Bartolo:2010bu, Seery:2010kh,
Kahya:2010xh}. (See also the recent discussions in
Refs.~\cite{Polyakov:2009nq, Marolf:2010zp, Hollands:2010pr, Rajaraman:2010zx,
Rajaraman:2010xd, Krotov:2010ma, Marolf:2011sh, Kitamoto:2010si, Kitamoto:2010et}.) In our previous works~\cite{IRgauge_L, IRgauge}, we
showed that in single field models of inflation the IR
divergence is an unphysical artifact, which disappears in requesting the genuine
gauge-invariance. Namely, initial quantum states are requested to
satisfy several conditions in order to respect the gauge invariance in the local
universe. The importance of the genuine gauge invariance is recognized
as well in the study of the primordial non-Gaussianity~\cite{IRNG}. It
is remarkable that the tree-level bi-spectrum calculated in the conventional perturbation theory
vanishes in the squeezed limit under the request of the genuine gauge invariance. This fact
emphasizes the importance to investigate primordial fluctuations based
on the genuine gauge invariant perturbation theory to yield the
theoretical prediction of fluctuations, which are to be compared with
the observations.

These arguments have been so far elaborated in single field models of
inflation. In this paper, we extend our argument to multi-field models. In
multi-field models, the issue of gauge-invariance becomes more delicate
because of the presence of the entropy field. If the entropy field is
massless, the loop correction of the entropy field diverges due to IR
contributions. In contrast to the IR divergence from the adiabatic 
field, the IR divergence from the entropy field is conceived to be
irrelevant to the presence of the gauge degrees of freedom~\cite{IRmulti}. The purpose
of this paper is not to provide ways of regularization, but to
provide one way to realize the genuine gauge invariance in the presence
of the entropy field. For this purpose, we need to discriminate between the IR divergences
from the gauge effects and those from different origins. In this paper
we consider two-field models of inflation, but an extension to inflation
models with more than two fields would proceed straightforwardly.

Our paper is organized as follows. In Sec.~\ref{Sec:Review}, we give the
setup of our problem and after that we briefly review a method to calculate
genuinely gauge-invariant quantities. In Sec.~\ref{Sec:Decoupled}, we
study the implications of the genuine gauge-invariant perturbation in
the two-field models with the pure adiabatic and entropy fields. After
we derive the gauge-invariance condition, we
calculate the genuinely gauge-invariant bi-spectrum in this model. In
Sec.~\ref{Sec:Coupled}, we extend our arguments to more general
two-field models where the adiabatic and entropy fields are coupled even
at liner order. Our results are summarized in Sec.~\ref{Sec:Conclusion}.

\section{Genuine gauge-invariant perturbations}
\label{Sec:Review}
In this section, we first give the set up of the problem. After that, we
provide one way to perform the genuinely gauge-invariant perturbation
theory, following the discussion in our previous works~\cite{IRgauge_L, 
IRgauge}.    

\subsection{Basic equations} 
\label{SSec:Basiceq}
We consider the two-field inflation models with the standard kinetic term
whose action take the form  
\begin{eqnarray}
 S &\!\!=\!\!& \frac{\Mp^2}{2} \!\int\! \sqrt{-g}  \left[R -  g^{\mu\nu}\phi^I_{,\mu}
  \phi_{I,\nu}
 - 2 V(\phi_1,\,\phi_2)  \right] \dd^4x, \nonumber \\
\end{eqnarray}
where $\Mp$ is the Planck mass and multiplying $1/\Mp$ the scalar fields
were rescaled to be dimensionless. The field-space metric for $\phi^I$
is assumed to be given by the $2\times2$ unit matrix.

The ADM formalism has been utilized 
to derive the action of the dynamical variables particularly in the
non-linear perturbation theory~\cite{Maldacena2002}. Using the decomposed metric
\begin{eqnarray}
 \dd s^2 = - N^2 \dd t^2  + h_{ij} (\dd x^i + N^i \dd t) (\dd x^j + N^j
  \dd t)~, 
\end{eqnarray}
the action is rewritten as
\begin{eqnarray}
 S&\!=&\!\frac{\Mp^2}{2} \!\int\! \sqrt{h} \Bigl[\, N \,\sR - 2 N
  V(\phi_1,\,\phi_2 ) + \frac{1}{N} (E_{ij} E^{ij} - E^2) 
 \cr && \qquad \qquad \quad \,
 + \frac{1}{N}  ( \partial_t \phi_I
  - N^i \partial_i \phi_I ) ( \partial_t \phi^I
  - N^i \partial_i \phi^I )  \cr && \qquad  \qquad \quad \,
 - N h^{ij} \partial_i \phi_I
  \partial_j \phi^I \,\Bigr] \dd^4x~,  \label{Exp:SADM}
\end{eqnarray}
where $\sR$ is the three-dimensional scalar curvature and $E_{ij}$ and $E$ are defined by 
\begin{eqnarray}
 E_{ij} = \frac{1}{2} \left( \partial_t h_{ij} - D_i N_j
 - D_j N_i \right), \quad E = h^{ij} E_{ij} 
\end{eqnarray}
with the three-dimensional covariant derivative $D_i$ defined by
$h_{ij}$. The spatial indices are raised and lowered by $h_{ij}$.

Perturbing the scalar fields as $\phi_I+\delta \phi_I$,
the background equations of motion are derived as
\begin{align}
 & \ddot{\phi}_I + 3 \dot{\rho} \dot{\phi}_I +  \partial V/\partial
 \phi^I=0\,, \label{Eq:KGoriginal} \\
 & 6 \dot{\rho}^2 = \dot{\phi}_I \dot{\phi}^I + 2 V(\phi_1, \phi_2)\,
 \label{Eq:Foriginal},
\end{align}
where the dot denotes the derivative by the cosmological time. Using
these equations, we also have
\begin{eqnarray}
 \ddot{\rho} = - \frac{1}{2} \dot{\phi}_I \dot{\phi}^I\,.
\end{eqnarray}

\subsection{Gauge-invariant operator} \label{SSec:GIQ}
In this subsection, we describe our way to calculate the gauge-invariant
quantity in the local universe.
One simple way to preserve the gauge-invariance is to calculate
fluctuations in a completely fixed slicing and threading. The
time-slicing can be easily fixed by adapting the gauge condition at each
spacetime point. By contrast, the complete fixing of the spatial
coordinates is not easy-going in the local universe because of the
degrees of freedoms in non-normalizable transformations~\cite{IRgauge_L,
IRgauge}. We therefore seek for the way to construct the manifestly
invariant quantity under the change in the spatial coordinates instead of
fixing them completely.

We fix the time slicing by eliminating the fluctuation in the adiabatic field.
We also adapt the gauge conditions on the spatial coordinates, taking the
spatial metric as
\begin{align}
 & h_{ij} = e^{2(\rho+\zeta)} \left[ e^{\delta \gamma} \right]_{ij}\,, 
\end{align}
where $\delta \gamma_{ij}$ satisfies the transverse and traceless
conditions $\delta \gamma^i\!_i=\partial_i \delta \gamma^i\!_j=0$.
While these conditions do not fix the spatial coordinates completely in
the local universe, we need not worry about the presence of the residual
gauge modes only if we consider the genuinely gauge-invariant quantity.
We can construct the genuinely gauge-invariant operator by making use of the scalar quantities
under the three dimensional spatial diffeomorphism. Following our previous works~\cite{IRgauge_L,
IRgauge}, we pick up the scalar curvature $\sR$ as such a scalar quantity. 
Although scalar quantities also vary under
the gauge transformation because of the change of their arguments
$x^i$, this gauge ambiguity does not appear in the $n$-point functions
of these quantities with the arguments specified in a gauge-invariant
manner. In order to specify the arguments of the $n$-point functions
in a gauge invariant manner, we measure distances from an
arbitrary reference point to the $n$ vertices by means of 
the geodesic distance obtained by solving the spatial three-dimensional 
geodesic equation: 
\begin{eqnarray}
 \frac{\dd^2 x^i}{\dd \lambda^2} +  {^s \Gamma^i}_{jk} \frac{\dd
  x^j}{\dd \lambda} \frac{\dd x^k}{\dd \lambda} =0~,
\label{GE}
\end{eqnarray}
where ${^s \Gamma^i}_{jk}$ is the Christoffel symbol with respect to 
the three dimensional spatial metric on a constant time hypersurface and
$\lambda$ is the affine parameter. 
We consider the three-dimensional geodesics whose affine parameter ranges
from $\lambda=0$ to $1$ with the initial ``velocity'' given by
\begin{eqnarray}
 \left.{\dd x^i(\bm{X},\lambda)
\over \dd \lambda}\right\vert_{\lambda=0}\!= e^{-\zeta(\lambda=0)} \left[ 
 e^{-\delta \gamma(\lambda=0)/2}\right]^i\!_j 
X^j\,. \label{IC}
\end{eqnarray}
We identify a point in the geodesic normal coordinates $X^i$ 
with the end point of the geodesic, $x^i(\bm{X},\lambda=1)$. 
Expanding the global coordinates $x^i$ by the geodesic normal
coordinates $X^i$ as $x^i(\bm{X}) =: X^i + \delta x^i(\bm{X})$, a genuine gauge invariant
variable can be given by
\begin{align}
 {^g\!R}(X) &:= \sR (t,\, x^i(\bm{X})) \cr
 & = \sum_{n=0}^\infty \frac{\delta x^{i_1} \cdots \delta x^{i_n}\!}{n!}
 \partial_{X^{i_1}}\! \cdots \partial_{X^{i_n}}{^s\!R}(t,\,
 X^i), \cr \label{Def:gR}
\end{align}
where we introduced the abbreviated notation $X:=\{t,\,\bm{X}\}$. 
Similar quantities to our genuine gauge-invariant variable were studied
in Refs.~\cite{Giddings:2010nc, Byrnes:2010yc, Gerstenlauer:2011ti,
Giddings:2011zd, Chialva:2011bg}, while the authors of
Refs.~\cite{Giddings:2010nc, Byrnes:2010yc, Gerstenlauer:2011ti,
Giddings:2011zd} seem to have the different opinions regarding what are actual
observables.

In the following section, we use the equality $\Approx$ which is valid
only when we neglect the terms which do not yield IR divergences
in the one-loop corrections to $\langle \gR \gR \rangle$. When we write
down the Heisenberg operator in terms of the interaction picture fields
for the adiabatic and entropy fields, this is equivalent to keep only
terms without spatial/time derivatives and the terms which include only
one interaction picture field with differentiation. In deriving the
gauge-invariance condition from the IR regularity of 
$\langle \gR \gR \rangle$, we neglect the possibly divergent loop
corrections whose loop integrals are composed only
of the entropy field, because they are irrelevant to the gauge effects.
Using the equality $\Approx$, the geodesic normal coordinates $X^i$ are related to the global coordinates
$x^i$ as
\begin{align}
 &x^i(\bm{X}) \Approx 
 e^{-\zeta} \left[ e^{-\delta \gamma/2} \right]^i\!_j X^j\,.
 \label{Rel:xX}
\end{align}

Abbreviating the unimportant
prefactor, we simply denote the scalar curvature $\sR$ as
\begin{align}
 & \sR \Approx e^{-2\zeta} \left[ e^{- \delta \gamma} \right]^{ij} \partial_{ij} \zeta\,.
\end{align}
The gravitational wave field $\delta \gamma_{ij}$ can also generate the IR
divergent with 
$\langle \delta \gamma_{ij} \delta \gamma_{kl} \rangle$. 
We can easily confirm that these contributions are generated only through
the interaction between the adiabatic field and the gravitational wave
field and that the entropy field does not produce another potentially
divergent terms with the gravitational wave field. Therefore, repeating the
same argument as in Ref.~\cite{IRgauge}, we can show that the
contributions from the gravitational waves are canceled in the
gauge-invariant operator $\gR$. To avoid the repetition of calculations,
we here do not explicitly write down the contributions from the
gravitational wave field.

Expanding $\zeta$ as $\zeta=\zeta_1 + \zeta_2 + \zeta_3 + ...$ where we
denote $\zeta_1$ as $\psi := \zeta_1$, the spatial curvature $\sR$ is
expressed as 
\begin{align}
 & \sR_1 = \partial^2 \psi\,, \qquad 
 \sR_2 \Approx \partial^2 \zeta_2 - 2 \psi \partial^2 \psi\,, \cr
 & \sR_3 \Approx \partial^2 \zeta_3 - 2(\zeta_2 \partial^2 \psi + \psi
 \partial^2 \zeta_2) + 2 \psi^2 \partial^2 \psi\,. \label{Exp:sR}
\end{align}
Using Eq.~(\ref{Rel:xX}), the difference between the geodesic normal
coordinates and the global ones is expanded as
\begin{align}
 & \delta x^i= \delta x^i_1+ \delta x^i_2 + \cdots\,
\end{align}
where
\begin{align}
 \delta x^i_1 &\Approx - \psi X^i, \qquad
 \delta x^i_2 \Approx - \zeta_2 X^i  + \frac{1}{2} \psi^2 X^i ~.  \label{Exp:dx}
\end{align}
Using Eqs.~(\ref{Def:gR}), (\ref{Exp:sR}), and (\ref{Exp:dx}), we have
\begin{align}
 \gR_1 &= \partial^2 \psi\,, \label{Exp:gR1} \\
 \gR_2 &\Approx \partial^2
 \zeta_2 - \psi \partial^2 X^i \partial_{X^i} \psi\,, \label{Exp:gR2} \\
 \gR_3 &\Approx \partial^2 \zeta_3 - \zeta_2 \partial^2 (X^i \partial_{X^i} ) \psi
 \nonumber \\
 & \qquad \quad - \psi \partial^2 (X^i \partial_{X^i}) \zeta_2 
  + \frac{1}{2} \psi^2
 \partial^2 (X^i \partial_{X^i})^2 \psi\,.  \label{Exp:gR3}
\end{align}

\section{Models with the pure entropy field}  \label{Sec:Decoupled}
In this and next sections, we investigate the predictions of the
genuine gauge-invariant perturbation theory in two kinds of models. In
general, the adiabatic and entropy fields can be coupled even in liner
order. We first study a particular model where
these two-fields are decoupled at linear order,
deferring the study of coupled models to the next section.

\subsection{Non-linear perturbations}
In this section, we consider the two-field model where the field
$\phi_1$ dominates the background.
We use the horizon flow functions~\cite{Schwarz:2001vv} defined in a similar way to those in single
field models as
\begin{align}
 & \varepsilon_1:= \frac{\dot{\phi}_1^2}{2\dot{\rho}^2} , \qquad
 \varepsilon_{m+1} := \frac{1}{\varepsilon_m} \frac{\dd
 \varepsilon_m}{\dd \rho}~\qquad  {\rm for}~m \geq 1.
 \label{Def:HFF} 
\end{align} 
Assuming that the horizon flow functions $\varepsilon_m$ are
all small of ${\cal O}(\varepsilon)$, we neglect the terms of ${\cal O}(\varepsilon^3)$.
We also employ the assumption:
\begin{align}
 \frac{\dot{\phi_2}}{\dot{\phi}_1} = {\cal O} (\varepsilon^2)\,.
 \label{Cond:D}
\end{align}
This condition requests that, within our approximation, the field $\phi_1$ and
$\phi_2$ become the pure adiabatic and entropy fields, respectively. 
We define the gauge-invariant quantity $\gR$ on the slicing:
\begin{equation}
 \delta\phi_1=0~.
\end{equation}

To evaluate the genuine gauge-invariant variable, it is convenient to
perform the calculation in the flat gauge:
\begin{eqnarray}
 \tilde{h}_{ij}
  =e^{2\rho} \left[ e^{\delta \tilde{\gamma}} \right]_{ij}, \quad
 {\delta \tilde\gamma^i}_i=0= \partial_i \delta {{\tilde\gamma}^i}_j, \label{Exp:metricflat}
\end{eqnarray}
where all the interaction vertexes are explicitly suppressed by the
slow-roll parameters~\cite{Maldacena2002, IRsingle}. We associate a
tilde with the metric perturbations in the flat gauge to discriminate
those in the comoving gauge. Here again we can neglect the contributions from the gravitational wave
field, which do not appear in $\gR$.
The action in this gauge is given by
\begin{align}
 S&\Approx \frac{\Mp^2}{2} \!\!\int\!\dd t\, \dd^3 \bm{x} e^{3\rho}
 \Bigl[ \tilde{N}^{-1} ( - 6 \dot{\rho}^2 + 4 \dot{\rho} \partial_i
 \tilde{N}^i )  
   \cr & \qquad + \tilde{N}^{-1}\!\! 
 \left( \dot\phi_I + \dot\varphi_I - \tilde{N}^i
 \partial_i \varphi_I
 \right)\! \left( \dot\phi^I + \dot\varphi^I - \tilde{N}^i
 \partial_i \varphi^I
 \right) \cr & \qquad 
 - 2 \tilde{N} \!\sum_{m=0}^\infty \frac{V_{I_1 \cdots I_m}}{m!}
  \varphi^{I_1}\!\! \cdots \varphi^{I_m}
 - \tilde{N}  \tilde{h}^{ij} \partial_i
  \varphi_I \partial_j  \varphi^I  \Bigr]\!, \cr
 \label{Exp:S/flat}
\end{align}
where we introduced
$V_{I_1\cdots I_n}:= \partial^n V/\partial \phi^{I_1}\cdots \partial\phi^{I_n}$
and $\delta \phi_I := \varphi_I$. Using the background equations and
Eq.~(\ref{Cond:D}), the derivatives of the potential are quantified as
\begin{align}
 & \frac{V_2}{\dot{\rho} \dot{\phi}_1} = {\cal O}(\varepsilon^2)\,,\quad
 \frac{V_{12}}{\dot{\rho}^2} = {\cal
 O}(\varepsilon^3)\,, \quad
 \frac{V_{112}}{\dot{\rho} \dot{\phi}_1} = {\cal O}(\varepsilon^2)\,.
\end{align}

The curvature perturbation in the gauge $\delta \phi_1=0$
is related to the fluctuation of the dimensionless scalar field (divided
by $\Mp$) in the flat gauge $\varphi_1$ as
\begin{align}
 \zeta& \Approx \zeta_n
   +\zeta_n\partial_\rho\zeta_n 
   +\frac{\varepsilon_2}{4} \zeta_n^2
  +{\zeta_n^2 \partial_\rho^2\zeta_n\over 2} \cr & \qquad \qquad 
  +{3 \varepsilon_2\zeta_n^2 \partial_\rho \zeta_n\over 4} 
  + \frac{1}{12} \varepsilon_2 (\varepsilon_2 +
 2\varepsilon_3)\zeta_n^3, \label{Exp:zeta0}
\end{align}   
where we have introduced
\begin{align}
 & \zeta_n:=- (\dot\rho/ \dot\phi_1)\, \varphi_1\,.
\end{align}

Variation of the total action with respect to $\varphi_I$
yields 
\begin{align}
 & e^{-3\rho}  \partial_t
 \left[ \frac{e^{3\rho}}{\tilde{N}} \left(\dot\phi_I + \dot\varphi_I \right)
 \right] + \tilde{N}\! \sum_{m=0} \frac{V_{I I_1\cdots I_m}}{m!}
 \varphi^{I_1} \cdots \varphi^{I_m} \nonumber \\
 & \quad \quad - \left( \dot{\phi}_I + \dot{\varphi}_I \right) \frac{1}{\tilde{N}}
 \partial_i \tilde{N}^i - \tilde{N} e^{-2\rho}
 \partial^2 \varphi_I  \Approx 0~.  \label{Eq:flat}
\end{align}
Variations with respect to the lapse function and the shift vector yield the
Hamiltonian constraint:
\begin{align}
 &(\tilde{N}^2-1) V  +  \tilde{N}^2 \sum_{m=1} \! \frac{V_{I_1\cdots I_m}}{m!}
 \varphi^{I_1}\cdots\varphi^{I_m} \cr & \qquad \qquad 
  + 2 \dot\rho \partial_i \tilde{N}^i +  \dot{\phi}_I \dot{\varphi}^I +
 \frac{1}{2}  \dot\varphi_I \dot{\varphi}^I \Approx 0\,,
\end{align}
and the momentum constraints:
\begin{eqnarray}
 2 \dot{\rho} \partial_i \tilde{N} - 
  \tilde{N}( \dot{\phi}_I \partial_i \varphi^I + \partial_i \varphi_I
  \dot{\varphi}^I) \Approx 0\,.
\end{eqnarray}
From the form of the constraint equations, we can confirm the presence
of the degrees of freedom which appear from boundary
conditions in solving these equations.
Repeating a similar analysis to the one in Ref.~\cite{IRgauge_L}, these
degrees of freedom are found to be the residual gauge degrees of freedom,
which can change the average value of $\zeta_n$ at each order in perturbation.

These constraint equations are solved to give
\begin{eqnarray}
  \delta \tilde{N}
  & \Approx&  -\varepsilon_1 \zeta_n+ \frac{\varepsilon_1^2}{2} 
  \zeta_{n}^2 +\frac{1}{4} \varepsilon_1 \varepsilon_2 \left( \zeta_n^2
							+ \sigma_n^2
						       \right),
  \label{Exp:N/flat} \\
  \frac{1}{\dot\rho} \partial_i \tilde{N}^i& \Approx& \varepsilon_1
  \partial_\rho \zeta_n
  -\frac{1}{2} \varepsilon_1 \varepsilon_2
  \left( \zeta_n \partial_\rho \zeta_n + \sigma_n \partial_\rho \sigma_n
  \right) \cr
 && \qquad  - \frac{1}{4} \varepsilon_1 \left(2 \eta_{22} +
					     3 \varepsilon_2
					    \right) \sigma_n^2\,, \label{Exp:Ni/flat}
\end{eqnarray}
where, in a similar manner to $\zeta_n$, we introduced
\begin{align}
 & \sigma_n := - (\dot{\rho}/ \dot{\phi}_1) \, \varphi_2\,.
\end{align}
Substituting Eqs.~(\ref{Exp:N/flat}) and (\ref{Exp:Ni/flat}) into
Eq.~(\ref{Eq:flat}), the evolution equation of $\zeta_n$ is recast into 
a rather compact expression, 
\begin{eqnarray}
 {\cal L} \zeta_n&\Approx&  
 \left[ -2\varepsilon_1 \zeta_n + \frac{1}{2}
			      \varepsilon_1 (4\varepsilon_1 +
			      \varepsilon_2) \zeta_n^2 \right] \frac{e^{-2\rho}}{{\dot\rho}^2}
   \partial^2 \zeta_n 
\cr && \quad 
-  \varepsilon_1 \varepsilon_2  \zeta_n
  \partial_\rho \zeta_n - \frac{3}{4}  \varepsilon_2
  \varepsilon_3  \zeta_n^2\,, 
  \label{Eq:flat2}
\end{eqnarray} 
where the differential operator ${\cal L}$ is defined by 
\begin{equation}
{\cal L} :=  \partial_\rho^2 +  (3 - \varepsilon_1+ \varepsilon_2) \partial_\rho
 - \frac{e^{-2\rho}}{{\dot\rho}^2} \partial^2.
\end{equation}
Note that the adiabatic field $\zeta_n$ is decoupled from the entropy
field $\sigma_n$ at linear order as the result of requiring
Eq.~(\ref{Cond:D}). We also find that the interaction terms with $\zeta_n \sigma_n$
and $\zeta_n^2 \sigma_n$ are suppressed at the order of ${\cal O}(\varepsilon^3)$. 
Since the entropy field $s_n$ is not included in Eqs.~(\ref{Exp:zeta0})
and (\ref{Eq:flat2}), the curvature perturbation $\zeta$ and the
gauge-invariant spatial curvature $\gR$ are given in the same form as
those in single field models. Therefore, in the decoupled model which satisfies
Eq.~(\ref{Cond:D}), the entropy field $s_n$ does not contribute to the
IR divergence that originates from the gauge effect.

Reflecting these facts, the gauge-invariance conditions derived by
requesting the regularity of loop corrections from the adiabatic
field take the same form as those in the single field model. We briefly
describe this result, deferring the detailed explanation to
Ref.~\cite{IRgauge}. Here, we expand the interaction picture field $\psi$ as
\begin{align}
 & \psi(X)= \int \frac{\dd^3 \bm{k}}{(2\pi)^{3/2}}
 \left[\, \psi_{\sbm{k}}(X) a_{\sbm{k}} + \rm{h.c.}\, \right]
\end{align}
with the creation and annihilation operators that satisfy
\begin{align}
 & \left[ a_{\sbm{k}},\, a_{\sbm{k}'}^\dagger
 \right]= \delta^{(3)} (\bm{k}- \bm{k}')\,.
\end{align}
If the positive frequency function satisfy
\begin{align}
 & \Bigl[ \bigl\{1+ \varepsilon_1 ( 1+\varepsilon_1 +
   \varepsilon_2) \bigr\} \partial_\rho -  X^i \partial_{X^i}
 +\varepsilon_1 + \frac{1}{2} \varepsilon_2 + 2 \xi_2
  \cr & \qquad \qquad - \frac{3}{2}{\cal
 L}_k^{-1}\varepsilon_2 (2\varepsilon_1 + \varepsilon_2) \Bigr] \psi_{\sbm{k}}(X)
  = - D_k \psi_{\sbm{k}}(X)\,,  \label{Cond:GI}
\end{align}
where $D_k$ and ${\cal L}_k$ are defined as
\begin{align}
 & D_k := \bm{k}\cdot \partial_{\sbm{k}} + 3/2\,, \\
 & {\cal L}_k :=  \partial_\rho^2 +  (3 - \varepsilon_1+ \varepsilon_2) \partial_\rho
 + \frac{e^{-2\rho}}{{\dot\rho}^2} k^2\,,
\end{align}
the gauge invariant operator $\gR_2$ is compactly summed as
\begin{align}
 & \gR_2 \Approx -(1+\lambda_2) \partial^2 D_k \psi\,.
\end{align}
Here $\xi_2$ and $\lambda_2$ are time-dependent functions of ${\cal
 O}(\varepsilon^2)$ and they appeared as degrees of freedom in solving
 Eq.~(\ref{Eq:flat2}) at second order in perturbation. We can 
 also find similar degrees of freedom at third order in perturbation and
 we label them by time dependent functions  $\xi_3$ and $\lambda_3$ of
 ${\cal O}(\varepsilon^2)$. If we choose them correctly~\cite{IRgauge},
the gauge invariant operator $\gR_3$ is compactly rewritten as 
\begin{align}
 ^g\! R_3& \Approx {1\over 2}\psi^2 \partial^2\!\left[ (1+ 2\lambda_2)
    D_k^2\psi -\mu D_k\psi  \right] - \delta \zeta_{n,2} \partial^2 D_k \psi \,,
\end{align}
where we defined $\delta \zeta_{n,2}$ and $\mu$ as
\begin{align}
 & \delta \zeta_{n,2} := - {\cal L}^{-1} \frac{3}{4} \varepsilon_2 (2\varepsilon_1 + \varepsilon_2) \psi^2\,,\\
 & \mu:=\varepsilon_1 + \frac{1}{2} \varepsilon_2-3\varepsilon_1^2+ \frac{1}{2} \varepsilon_1
 \varepsilon_2 + 2(\xi_2+\lambda_3)\,.
\end{align}
We finally find that the possibly divergent terms become the total
derivative form as
\begin{align}
 &  \langle \left\{  {}^g\!R(X_1),\, {}^g\!R(X_2) \right\} \rangle \cr
 &~ \Approx \frac{1}{2} \langle \psi^2 \rangle
\!\int\! {\dd(\log k) \over 2 \pi^2} \bigl\{ 
   (1+ 2 \lambda_2) \partial_{\log k}^2 
   - \mu  \partial_{\log k} 
   \bigr\}  \cr & \qquad \qquad \qquad \qquad \qquad  \times\!  \left\{ k^7 \psi_{\bk}(X_1)
 \psi_{\bk}^*(X_2) + \left( {\rm c.c.} \right) \right\}
   \cr
 &  \quad - \langle \delta \zeta_{n,2} \rangle \!\int\!  {\dd(\log k) \over 2 \pi^2}
 \partial_{\log k}\! \left\{ k^7 \psi_{\bk}(X_1)
 \psi_{\bk}^*(X_2) + \left( {\rm c.c.} \right) \right\},\cr
 \label{Exp:2pgR} 
\end{align}
and hence they vanish. Here we symmetrized about $X_1$ and $X_2$. 
We thus obtain the gauge-invariance conditions by requesting the regularity of the loop corrections
from the adiabatic field. Note that the preservation of the genuine
gauge-invariance guarantees the regularity of the loop corrections from
the adiabatic field, but it does not guarantee the regularity of those
from the entropy field. Therefore, the divergent terms with 
$\langle \chi^2 \rangle$ are still left in the left-hand side of Eq.~(\ref{Exp:2pgR}) and they should
be regularized by another method.

\subsection{Primordial non-Gaussianity}
In this subsection, we study the primordial non-Gaussianity, requesting
the gauge-invariance conditions. In calculating the tree-level
bi-spectrum, we can construct the genuine gauge-invariant variable using
the curvature perturbation $\zeta$ in stead of the spatial scalar
curvature $\sR$. If we consider the
modes with $k \gg 1/L_{\rm obs}$, where $L_{\rm obs}$ denotes the
observable scale in the comoving coordinates, the curvature
perturbation $\zeta$ evaluated in the geodesic normal coordinates:
\begin{align}
 \gz(X) &:= \zeta (\rho,\,x^i(\bm{X})) \cr
 &=\zeta(X) + \delta x^i \partial_i \zeta |_{x^i=X^i}+\cdots\, 
\label{Def:gz} 
\end{align}
becomes the gauge-invariant operator at least up to the second order in
perturbation~\cite{IRNG}.

Introducing the Fourier modes of $\gz$ as
\begin{align}
 &\gz_{\sbm{k}}(\rho) = \int \frac{{\rm d}^3 \bm{X}}{(2\pi)^{3/2}}\, 
 e^{- i \sbm{k} \cdot \sbm{X}} \gz(\rho,\, \bm{X})\,, 
\end{align}
we calculate the bispectrum of $\gz$s at the leading order in perturbation.
Expanding $\gz$ as $\gz_{\sbm{k}}= \psi_{\sbm{k}} + \gz_{\sbm{k},2}+ \cdots$, we have
\begin{align}
 &\langle \gz_{\bk_1} \gz_{\bk_2} \gz_{\bk_3} \rangle \nonumber\\
 & = 
  \langle \psi_{\bk_1} \psi_{\bk_2} \gz_{\bk_3,2}  \rangle + 
  \langle \psi_{\bk_1}  \gz_{\bk_2,2} \psi_{\bk_3}  \rangle +
   \langle  \gz_{\bk_1,2} \psi_{\bk_2} \psi_{\bk_3}  \rangle\,. \label{Exp:3point}
\end{align}
In the calculation of the bi-spectrum for $\gz$, we again use the
equality $\Approx$, which picks up gauge-dependent terms. Using
Eqs.~(\ref{Exp:dx}) and (\ref{Exp:zeta0}), the gauge-invariant curvature
perturbation is expressed as
\begin{eqnarray}
 \gz \Approx \zeta_n
  + \zeta_n \left( \partial_\rho - X^i \partial_{X^i} \right) \zeta_n
  + \frac{1}{4} \varepsilon_2 \zeta_n^2\,, \label{Exp:zetag2}
\end{eqnarray}
where $\zeta_n$ is given by solving Eq.~(\ref{Eq:flat2}). Since there is
no contribution from the entropy field in Eqs.~(\ref{Eq:flat2}) and
(\ref{Exp:zetag2}), $\gz$ takes the same form as in single field models. 
As expected from these facts, repeating the same calculation as in
Ref.~\cite{IRNG}, we again find that, in the squeezed limit with 
$k_1 \ll k_2 \approx k_3$, the bi-spectrum vanishes as
\begin{align}
 & \langle \gz_{\sbm{k}_1} \gz_{\sbm{k}_2} \gz_{\sbm{k}_3} \rangle
 \Approx 0\,. \label{Exp:3pointR}
\end{align}
The bi-spectrum calculated in the conventional perturbation theory is
thus found to be dominated by the gauge artifact also in the two-field
model which satisfies the condition (\ref{Cond:D}). Note that, while the
three-point function generated from the cubic interaction $\zeta_n^3$ is
suppressed from the request of the genuine gauge invariance, the one
generated from the interaction 
$\zeta_n \sigma^2_n$ is not suppressed at all. Therefore, our
gauge-invariance conditions do not affect on the three-point function with one
adiabatic field and two entropy fields.

One important remark is in order regarding the applicability of our
argument. Precisely speaking, the expression of $\delta x^i$ given in Eq.~(\ref{Exp:dx})
can be reliably used only for the modes with $k \ll 1/L_{\rm obs}$. We
can however show that the additional terms which appear for $k \agt
1/L_{\rm obs}$ vanish separately in the squeezed limit~\cite{IRgauge}. Thus,
we can confirm that the result given in Eq.~(\ref{Exp:3pointR}) is robust
as long as we consider the super-horizon modes with $k \ll e^{\rho}H$.
During inflation, the scales relevant to the current observations go
far outside of the inflationary horizon, which leads to the natural assumption that
the observable scale is much larger than the horizon
scale $1/H$. Therefore, our result is applicable to the
observable modes which satisfy $1/L_{\rm obs} \ll k \ll e^\rho H$.

\section{General extension}  \label{Sec:Coupled}
In the previous section, we studied the two-field model of inflation
where the adiabatic and entropy fields decouple at linear order in
perturbation. In this model, the allowable form of interactions is
significantly restricted so that the entropy field does not contribute
to the potentially divergent terms with the loop integrals of the
adiabatic field. The predictions of the genuinely gauge-invariant
perturbation are then found to be essentially same as those in single field models of
inflation. In this section, we consider more general two-field models
where the adiabatic and entropy fields are coupled even at linear order.
The study of these models would help us to understand how the entropy
field contributes to the IR corrections affected by non-local gauge degrees
of freedom.

\subsection{Adiabatic and entropy perturbations}
In multi-field models, we need to care about the fact that the IR modes include the
physical degrees of freedom as well as the gauge degrees of freedom. 
The discrimination of the later from the former proceeds easily in the
model with the pure adiabatic and entropy fields, but in general it
becomes more complicated. To ease this procedure, we decompose the
fluctuations in the two fields into the horizontal and orthogonal
directions to the background trajectory as  
\begin{align}
 \left(
\begin{array}{c}
 \delta \sigma \\ \delta s  
\end{array}
 \right) = \frac{1}{\dot{\sigma}} \left(
\begin{array}{cc}
 \dot{\phi}_1 &  \dot{\phi}_2 \\[0.5pt]
 -  \dot{\phi}_2 & \dot{\phi}_1 \\ 
\end{array}
 \right) \left(
\begin{array}{c}
 \delta \phi_1 \\ \delta \phi_2  
\end{array}
 \right)  =: \Theta  \left(
\begin{array}{c}
 \delta \phi_1 \\ \delta \phi_2  
\end{array}
 \right), \label{Def:Theta}
\end{align}
where we introduced the $2\times 2$ time-dependent rotational matrix $\Theta$ with
$\dot{\sigma}$ defined as
$$
 \dot{\sigma} := \sqrt{(\dot{\phi}_1)^2 + (\dot{\phi}_2)^2}\,. 
$$
In this expression, the action (\ref{Exp:SADM}) is rewritten as
\begin{align}
 & S = \frac{\Mp^2}{2} \!\!\int\!\! \dd^4 x\, \sqrt{h} \biggl[ N \sR - 2 N
 V(\sigma,\, s) +  \frac{1}{N} \left( E^{ij} E_{ij} - E^2 \right)
 \nonumber \\ & \qquad \quad 
 +\! \frac{1}{N} \left( \dot{\sigma}+ \delta \dot{\sigma} - N^i \partial_i \delta \sigma
 - \dot{\theta} \delta s \right)^2 - N h^{ij} \partial_i \delta \sigma \partial_j \delta \sigma
 \nonumber \\  & \qquad \quad 
  +\! \frac{1}{N} \left(  \delta \dot{s} - N^i \partial_i \delta s
 + \dot{\theta} \delta \sigma \right)^2 - N h^{ij} \partial_i \delta s \partial_j \delta s
  \biggr]\,, 
\end{align}
where $\theta$ is the local rotation angle given by
$$
\theta:= \tan^{-1} ( \dot{\phi}_2 / \dot{\phi}_1 )\,.
$$
The model studied in the previous section corresponds to the particular
case with $\theta=\,$const, where the background trajectory is not
curved.

Using $\delta \sigma$ and $\delta s$, we rewrote the perturbation
expansion of the potential as
\begin{align*}
 & V(\sigma,\, s) = V_0 + V_\sigma \delta \sigma + V_s
 \delta s \cr & \qquad \qquad \quad  + \frac{1}{2} V_{\sigma \sigma} \delta \sigma^2 + V_{\sigma s}
 \delta \sigma \delta s + \frac{1}{2} V_{ss} \delta s^2  + \cdots\,,
\end{align*}
where $V_0$ denotes the background value of the potential. Note that
$V_{\alpha_1 .. \alpha_i .. \alpha_n}$ where 
$\alpha_i = \sigma,\,s$ are not the derivatives of the potential
in terms of $\sigma$ and $s$, but they are given by the linear
combinations of $V_{I_1 ..I_n}$. In general multi-field models, the
cross correlation between the adiabatic and entropy fields 
$\langle \delta \sigma  \delta s \rangle$ does not necessarily vanish. The loop
corrections with $\langle \delta \sigma  \delta s \rangle$ can also yield
divergent terms which are possibly affected by gauge degrees of
freedom. Keeping this in mind, here we consider the terms that can contribute to the
loop corrections with $\langle \delta \sigma^2 \rangle$ and also
$\langle \delta \sigma  \delta s \rangle$. As in the previous section, we keep terms
that include at most one interaction picture field for $\delta \sigma$ or $\delta s$ with differentiation,
but among them we neglect terms that are expanded only in terms of the
interaction picture field for $\delta s$.  In the
following we use the same equality $\Approx$ to denote a equality which
is valid in neglecting these terms.

We define the gauge-invariant scalar curvature $\gR$ on the time-slice
fixed by the gauge condition:
$$\delta \sigma=0.$$
We again change the gauge into the flat gauge given by
Eq.~(\ref{Exp:metricflat}). Noticing
the fact that the background evolution is characterized only by the
adiabatic field, we introduce the horizon-flow functions as
\begin{align}
 & \varepsilon_1:= \frac{\dot{\sigma}^2}{2\dot{\rho}^2} , \qquad
 \varepsilon_{m+1} := \frac{1}{\varepsilon_m} \frac{\dd
 \varepsilon_m}{\dd \rho}~\qquad  {\rm for}~m \geq 1.
 \label{Def:HFFc} 
\end{align}
In the following,  we assume that the change in the background trajectory takes place
satisfying $\dd \theta/\dd \rho = {\cal O}(\varepsilon)$. We also assume
that the mass of the entropy field satisfies $V_{ss}/\dot{\rho}^2={\cal O}(\varepsilon)$ so that
the entropy field participates in the generation of the fluctuation. In
the flat slicing, the action is given by
\begin{align}
 & S \Approx \frac{\Mp^2}{2}  \!\int\! \dd t\, \dd^3 {\bm x} e^{3\rho}
 \biggl[  \tilde{N}^{-1} (\dot\sigma + \delta \dot\sigma - \tilde{N}^i
 \partial_i \delta \sigma - \dot\theta \delta s)^2 \nonumber \\[0.5pt]
 & \qquad  + \tilde{N}^{-1} (\delta \dot{s} - \tilde{N}^i
 \partial_i \delta s + \dot\theta \delta \sigma)^2 
 - \tilde{N}
 \tilde{h}^{ij} \partial_i \delta \sigma \partial_j \delta \sigma 
\cr   & \qquad
- \tilde{N}
 \tilde{h}^{ij} \partial_i \delta s \partial_j \delta s 
+ \tilde{N}^{-1}(-6 {\dot\rho}^2 + 4 \dot\rho \partial_i
 \tilde{N}^i ) - 2 \tilde{N} V \biggr]\,.
\end{align}
The constraint equations are derived as
\begin{align}
 &  2 \dot\rho \partial_i \tilde{N}^i + \dot\sigma \delta \dot\sigma + \frac{1}{2}
 (\delta \dot\sigma - \dot\theta \delta s)^2 + \frac{1}{2} (\delta \dot{s} +
 \dot\theta \delta \sigma)^2 \cr
 & \quad + (\tilde{N}^2-1) V_0 + V_\sigma \delta \sigma \cr
 & \quad 
 +  \frac{1}{2} (V_{\sigma \sigma} \delta \sigma^2 + 2 V_{\sigma s}
 \delta \sigma \delta s
 + V_{ss} \delta s^2) + \cdots   \Approx 0\,,  \label{HconstC} 
\end{align}
and 
\begin{align}
 &  2 \dot\rho \partial_i \tilde{N} - \tilde{N} (\dot\sigma - \dot\theta
 \delta s) \partial_i \delta \sigma  - \tilde{N} \dot\theta \delta \sigma
 \partial_i \delta s \cr
 & \qquad \qquad \qquad - 2 \tilde{N} ( \delta \dot\sigma \partial_i \delta \sigma + 
 \delta \dot{s} \partial_i \delta s ) \Approx 0\,, \label{MconstC}
\end{align}
where, in Eq.~(\ref{HconstC}), we abbreviated the higher-order terms in $V(\sigma,\, s)$. In
deriving the Hamiltonian constraint (\ref{HconstC}), we used 
$V_s = - \dot\sigma \dot\theta$, which is given by varying the  action
with respect to $\delta s$. Consulting Eqs.~(\ref{HconstC}) and
(\ref{MconstC}), we again find that the degrees of freedom in the boundary
conditions are intrinsically attributed to the gauge degrees of freedom in the
adiabatic fluctuation, while the entropy fluctuations are, at non-linear
order, also affected by the gauge modes through the change in the
lower-order adiabatic fluctuation. Varying the action with respect to
$\delta \sigma$, we have the equation of motion for $\delta \sigma$ as
\begin{align}
 &e^{-3\rho}\partial_t \left\{ \frac{e^{3\rho}}{\tilde{N}} (\dot\sigma + \delta \dot\sigma - \dot\theta
 \delta s) \right\} -  (\dot\sigma + \delta \dot\sigma - \dot\theta
 \delta s) \frac{\partial_i \tilde{N}^i}{\tilde{N}} \cr 
 & \qquad + \tilde{N} \Bigl(V_\sigma + V_{\sigma \sigma} \delta
 \sigma + V_{\sigma s} \delta s \cr
 & \qquad \qquad \qquad \quad  + \frac{1}{2} V_{\sigma\sigma\sigma}
 \delta \sigma^2 + \frac{1}{2}  V_{\sigma s s} \delta s^2 + V_{\sigma\sigma s}
 \delta \sigma \delta s \Bigr)  \nonumber \\
 & \qquad - \frac{\dot\theta}{\tilde{N}} (\delta \dot{s} + \dot\theta \delta \sigma)
 - \tilde{N}  e^{-2\rho} \partial^2 \delta \sigma \Approx 0\,. \label{Eq:sigma}
\end{align}

From a similar calculation to that in Appendix A
of Ref.~\cite{IRgauge}, the relation between the fluctuations in these two
gauges is obtained as
\begin{align}
 & \zeta \Approx \zeta_n + \zeta_n \partial_\rho \zeta_n + \frac{1}{4}
 \varepsilon_2 \zeta_n^2 - \frac{\dd \theta}{\dd \rho} \zeta_n s_n\cr
 & \qquad  + \frac{1}{2} \zeta_n^2 \partial_\rho^2 \zeta_n + \frac{3}{4}
 \varepsilon_2 \zeta_n^2 \partial_\rho \zeta_n + \frac{1}{12}
 \varepsilon_2 (\varepsilon_2+2 \varepsilon_3) \zeta_n^3 \cr
 & \qquad - \frac{1}{3} \Bigl( \frac{\dd \theta}{\dd \rho}\Bigr)^2
 \zeta_n^3 - 2 \frac{\dd \theta}{\dd \rho} \zeta_n s_n \partial_\rho \zeta_n
 - \frac{\dd \theta}{\dd \rho} \zeta_n^2 \partial_\rho s_n \cr
 & \qquad -\frac{1}{2} \left( \frac{\dd^2 \theta}{\dd \rho^2} +
 \frac{3}{2} \varepsilon_2 \frac{\dd \theta}{\dd \rho} \right) \zeta_n^2
 s_n + \Bigl( \frac{\dd \theta}{\dd \rho}\Bigr)^2 \zeta_n s_n^2\,.
 \label{Rel:zetazetanC}
\end{align}
where we defined $\zeta_n$ and $s_n$ as
\begin{align}
 & \zeta_n:= - (\dot\rho/\dot\sigma) \delta \sigma\,, \qquad
 s_n:= - (\dot\rho/\dot\sigma) \delta s\,.
\end{align}

While in the case with $\dd \theta/\dd \rho \neq 0$ the evolution equations for $\zeta_n$ and $s_n$ are coupled even
at linear order, we can easily solve
Eq.~(\ref{Eq:sigma}) up to ${\cal O}(\varepsilon)$. In the following
calculations, we keep the terms up to this order, neglecting the terms
of ${\cal O}(\varepsilon^2)$. Then, the constraint
equations (\ref{HconstC}) and (\ref{MconstC}) are solved to give
\begin{align}
 & \delta \tilde{N} \Approx - \varepsilon_1 \zeta_n\,, \qquad
 \frac{1}{\dot{\rho}}  \partial_i \tilde{N}^i \Approx  \varepsilon_1
 \partial_\rho \zeta_n\,. \label{Exp:NNiC}
\end{align}
Using Eqs.~(\ref{Exp:NNiC}), the equation of motion for $\zeta_n$ is
recast into
\begin{align}
 & {\cal L} \zeta_n \Approx 2 \frac{\dd \theta}{\dd \rho} \partial_\rho
 s_n + \left( 3 \frac{\dd \theta}{\dd
 \rho} -  \frac{V_{\sigma s}}{{\dot\rho}^2} \right) s_n - 2
 \varepsilon_1 \zeta_n
 \frac{e^{-2\rho}}{{\dot\rho}^2} \partial^2  \zeta_n
\,, \label{Eq:eomc}
\end{align}
where the derivative operator ${\cal L}$ apparently takes the same form as the
single field case:
\begin{align}
 & {\cal L}:= \partial^2_\rho + (3-\varepsilon_1 + \varepsilon_2)
 \partial_\rho - \frac{e^{-2\rho}}{{\dot\rho}^2} \partial^2\,.
\end{align}
Up to this order, interaction terms with the entropy field do not appear
in the equation of motion for $\zeta_n$.
Allowing the introduction of homogeneous solutions, Eq.~(\ref{Eq:eomc})
is solved to give
\begin{align}
 & \zeta_n \Approx \psi + \varepsilon_1 \psi \partial_\rho \psi
 + \xi_2 \psi^2 + \xi_3 \psi^3  \cr & \qquad \qquad + 
 \mu_2 \psi \chi + \mu_3 \psi^2 \chi  + \hat{\mu}_2 \chi^2
 + \hat{\mu}_3 \psi \chi^2 \,, \label{Exp:zetanC}
\end{align}
where $\xi_i$, $\mu_i$, and $\hat{\mu}_i$ for $i=2,3$ are time-dependent
functions of ${\cal O}(\varepsilon)$. Noticing the fact that the
interaction picture field of $s_n$ satisfies
\begin{align}
 & {\cal L} \chi = {\cal O}(\varepsilon)\,,
\end{align}
we also included the homogeneous solutions with $\chi$.

\subsection{IR regularity and gauge-invariance conditions}
Now, we are ready to derive the gauge-invariance condition for the
coupled case. Using Eqs.~(\ref{Exp:gR2}) and (\ref{Exp:gR3}) together with
Eqs.~(\ref{Rel:zetazetanC}) and (\ref{Exp:zetanC}), we have 
\begin{align}
 &\gR_2 \Approx \psi \partial^2 \Bigl[ \Bigl\{
 (1+\varepsilon_1)\partial_\rho - X^i \partial_{X^i}
\, \cr &\qquad \qquad \qquad \qquad 
 + 2\xi_2 +
 \varepsilon_2/2 \Bigr\} \psi
+ \Bigl( \mu_2 - \frac{\dd \theta}{\dd \rho} \Bigr)
 \chi  \Bigr]\, \cr &\qquad \qquad
+ \chi \partial^2 \Bigl[ \Bigl( \mu_2 - \frac{\dd \theta}{\dd \rho} \Bigr)
 \psi + 2 \hat{\mu}_2 \chi \Bigr]\,, \label{Exp:gR2c}
\end{align}
and
\begin{align}
 & \gR_3 \Approx \psi^2 \partial^2 \biggl[ \frac{1}{2} (\partial_\rho - X^i
 \partial_{X^i})^2 \psi + 3 \xi_3 \psi + \mu_3 \chi \cr
 & \qquad \qquad \qquad\, + (\partial_\rho - X^i \partial_{X^i})\! \Bigl\{
 (\varepsilon_1 \partial_\rho + 3 \xi_2 + 3\varepsilon_2/4 ) \psi  \cr
 & \qquad \qquad \qquad \qquad \qquad \qquad  \qquad \quad +
  \Bigl(\mu_2- \frac{\dd \theta}{\dd \rho}  \Bigr) \chi \Bigr\} 
\biggr] \cr & \qquad + 2 \psi \chi
 \partial^2 \biggl[ \Bigl(\mu_2- \frac{\dd
 \theta}{\dd \rho}  \Bigr)  (\partial_\rho - X^i \partial_{X^i}) \psi \cr
 & \qquad \qquad \qquad \quad  + \hat{\mu}_2 (\partial_\rho - X^i \partial_{X^i})  \chi
 + \mu_3 \psi + \hat{\mu}_3 \chi 
 \biggr]\!. \cr \label{Exp:gR3c}
\end{align}
We keep the terms in the last line of Eq.~(\ref{Exp:gR2c}) and the terms
in the last two lines of Eq.~(\ref{Exp:gR3c}), which can yield the possibly divergent terms with
$\langle \psi \chi \rangle$.

Making use of the Gram-Schmidt normalization, we can prepare a set of
orthonormalized mode functions, which satisfy the
Klein-Gordon normalization. We expand 
$\psi^I:= - (\dot{\rho}/\dot{\sigma}) \delta \phi^I$ by the
orthonormalized mode function $\psi^I_{\alpha, \sbm{k}}$ where
$\alpha=1,2$ are the indices for the orthonormal bases.
Using these orthonormal bases, $\psi$ and $\chi$  are expanded as
\begin{align}
 & \psi(X) =  \sum_{\alpha=1}^2 \Theta_{1I}
 \psi^I_\alpha(X)\,, \label{Exp:psi} \\
 & \chi(X) =  \sum_{\alpha=1}^2 \Theta_{2I}
 \psi^I_\alpha(X)\,, \label{Exp:chi}
\end{align}
where the time-dependent matrix $\Theta$ is already given in
Eq.~(\ref{Def:Theta}) and we defined $\psi^I_\alpha(X)$ as
\begin{align}
 & \psi^I_\alpha(X)= \int \frac{\dd^3 \bm{k}}{(2\pi)^{3/2}}
 \left[\, \psi^I_{\alpha, \sbm{k}}(X) a_{\alpha,\,\sbm{k}} + \rm{h.c.}\, \right]
\end{align}
with the creation and annihilation operators that satisfy
\begin{align}
 & \left[ a_{\alpha,\, \sbm{k}},\, a_{\beta,\, \sbm{k}'}^\dagger
 \right]= \delta_{\alpha \beta} \delta^{(3)} (\bm{k}- \bm{k}')\,.
\end{align}
Substituting Eqs.~(\ref{Exp:psi}) and (\ref{Exp:chi}) into
Eqs.~(\ref{Exp:gR2c}) and (\ref{Exp:gR3c}), the gauge-invariant spatial
curvature is expanded as
\begin{align}
 & \gR_2 \Approx  \psi \sum_{\alpha=1}^2  \partial^2 \Bigl[ \Theta_{1I} \bigl\{ 
 (1+\varepsilon_1) \partial_\rho - X^i \partial_{X^i}
 \cr & \qquad \qquad \qquad \qquad \qquad + 2\xi_2 +
 \varepsilon_2/2 \bigr\} 
+ \Theta_{2I} \mu_2 \Bigr] \psi^I_\alpha
\cr & \qquad \, + 
\chi  \sum_{\alpha=1}^2 \partial^2 \Bigl[ \Bigl( \mu_2 - \frac{\dd \theta}{\dd \rho} \Bigr)
 \Theta_{1I} + 2 \hat{\mu}_2 \Theta_{2I}\Bigr] \psi^I_\alpha,
\end{align}
and
\begin{align}
 &\gR_3 \Approx  \psi^2  \sum_{\alpha=1}^2\partial^2 \biggl[\frac{1}{2} \Theta_{1I}
 (\partial_\rho - X^i \partial_{X^i})^2  \cr
 & \qquad \qquad \quad + \Theta_{1I}
 (3 \xi_2 + 3 \varepsilon_2/4 + \varepsilon_1 \partial_\rho)  (\partial_\rho - X^i \partial_{X^i})  \cr
 & \qquad \qquad \quad + \Theta_{2I}  \mu_2  (\partial_\rho - X^i
 \partial_{X^i}) + 3\xi_3 \Theta_{1I} + \mu_3 \Theta_{2I}
\biggr] \psi^I_\alpha  \cr
 & \qquad \quad +  2 \psi\chi \sum_{\alpha=1}^2 \partial^2 \biggl[
 \Theta_{1I} \Bigl(\mu_2- \frac{\dd
 \theta}{\dd \rho}  \Bigr)  (\partial_\rho - X^i \partial_{X^i})  \cr
 & \qquad \qquad \quad  + \Theta_{2I} \hat{\mu}_2  (\partial_\rho - X^i
 \partial_{X^i})
 + \mu_3 \Theta_{1I} + \hat{\mu}_3
 \Theta_{2I} \biggr] \psi^I_\alpha \,,\cr
\end{align}
where we noted 
$\dd \Theta_{1I}\!/\!\dd \rho= (\dd \theta\!/\!\dd \rho) \Theta_{2I}$.

In requesting the absence of the possibly divergent terms with 
$\langle \psi^2 \rangle$ and $\langle \psi \chi \rangle$, we obtain the gauge-invariance conditions as
\begin{align}
 &\, \Bigl[ \Theta_{1I} \bigl\{ 
 (1+\varepsilon_1) \partial_\rho - X^i \partial_{X^i}
+ 2\xi_2 +
 \varepsilon_2/2 \bigr\} 
+ \Theta_{2I} \mu_2 \Bigr] \psi^I_{\alpha,\, \sbm{k}} \nonumber \\
 &\, \qquad \qquad = - \Theta_{1I} D_k  \psi^I_{\alpha,\, \sbm{k}}
\,, \label{Exp:GCc1} \\
 &\,  \xi_3 = \mu_3=0\,, \label{Exp:GCc2} 
\end{align}
and 
\begin{align}
 & \mu_2 = \frac{\dd \theta}{\dd \rho}\,, \qquad
 \hat{\mu}_2=\hat{\mu}_3=0\,. \label{Exp:GCc3}
\end{align}
If all these conditions are satisfied, the gauge-invariant spatial
curvature $\gR$ is simply given by
\begin{align}
 & \gR_2 \Approx - \psi  \sum_{\alpha=1}^2 \partial^2 D_k \Theta_{1I} \psi^I_\alpha, \\
 & \gR_3 \Approx \frac{1}{2} \psi^2  \sum_{\alpha=1}^2 \partial^2 \biggl[ D_k^2 \Theta_{1I} - \Bigl(2\xi_2 +
 \frac{\varepsilon_2}{2} \Bigr) D_k \Theta_{1I} \biggr]\psi^I_\alpha\,.
\end{align}
Then, the potentially divergent terms in $\gR$ are found
to become the total derivative as
\begin{align}
 &  \langle \left\{  {}^g\!R(X_1),\, {}^g\!R(X_2) \right\} \rangle \cr
 &~ \Approx \frac{1}{2} \langle \psi^2 \rangle \Theta_{1I} \Theta_{1J}
\!\int\! {\dd(\log k) \over 2 \pi^2}\! \left\{ 
   \partial_{\log k}^2 
   - \left( 2\xi_2+\frac{\varepsilon_2}{2} \right)  \partial_{\log k} 
   \right\}  \cr & \qquad  \quad \times\! \sum_{\alpha=1}^2  \left\{ k^7
 \psi^I_{\alpha,\sbm{k}}(X_1) \psi^I_{\alpha,\sbm{k}}\!^*(X_2) 
  + (\rm{c.c.}) \right\},
\end{align}
and they vanish. We now understand that, after the choice of the appropriate initial
condition, the two-point function of the gauge-invariant operator $\gR$
is shown to be regular, leaving a possibly IR divergent
contribution with $\langle \chi^2 \rangle$, which is irrelevant to the
gauge effect.  

We requested Eq.~(\ref{Exp:GCc3}) to eliminate the possibly divergent
terms with $\langle \psi \chi \rangle$. However, at first glance it may
be unclear whether the realization of the gauge invariance truly requests the
absence of the terms with $\langle \psi \chi \rangle$ or not, because
the cross-correlation $\langle \psi \chi \rangle$ is also influenced by
the entropy field. To make this point clear, we note that 
$\langle \psi \chi \rangle$ no longer diverges if at least one of $\psi$
and $\chi$ is suppressed in the IR limit. If we remove the residual gauge
degrees of freedom, say by adapting the local gauge condition~\cite{IRsingle}, the adiabatic
field is suppressed in the IR limit so that the regularity of 
$\langle \psi \chi \rangle$ is sufficiently guaranteed. This indicates
that the cross-correlation $\langle \psi \chi \rangle$ does not diverge
in the genuine gauge-invariant quantities. Therefore, we can
request the absence of the IR divergence from 
$\langle \psi \chi \rangle$ as the necessary condition for the genuine gauge invariance.

Several remarks are in order regarding the gauge-invariance conditions
(\ref{Exp:GCc1})~-~(\ref{Exp:GCc3}). In harmony with the result in
single field models, the gauge invariance condition (\ref{Exp:GCc1})
almost uniquely determine the mode function for $\psi$
to that for the Bunch-Davies vacuum at the leading order in the slow-roll
approximation. This can be confirmed by following a similar argument to
the one in Refs.~\cite{IRgauge_L, IRgauge}. The gauge-invariance
conditions thus derived should be consistent with the equation of motion for the
adiabatic field. Using the Fourier expanded basis:
\begin{align}
 & \psi^I_{\alpha, \sbm{k}}(X) = 
 \frac{\dot{\rho}^2}{\dot{\phi}} \frac{1}{k^{3/2}} f^I_{\alpha, k}
 (\rho)\, e^{i \sbm{k} \cdot \sbm{X}}\,,
\end{align}
the gauge-invariance condition (\ref{Exp:GCc1}) can be recast into
\begin{align}
  \bigl\{ (1+\varepsilon_1) \partial_\rho + \bm{k} \cdot
 \partial_{\sbm{k}}
 + (2\xi_2 - \varepsilon_1) \bigr\} \Theta_{1I} f^I_\alpha =0\,, \label{Exp:GCc1v2}
\end{align}
where we also used the first equation in Eq.~(\ref{Exp:GCc3}).
From Eq.~(\ref{Eq:eomc}), we have the mode equation at liner order as
\begin{align}
 & \bar{\cal L}_k \Theta_{1I} f^I_{\alpha,k} 
-  \left( 3 \frac{\dd \theta}{\dd \rho}  - \frac{V_{\sigma
 s}}{\dot{\rho}^2} +2 \frac{\dd \theta}{\dd \rho}
 \partial_\rho \right) \Theta_{2I} f^I_{\alpha,k}=0, \label{Eq:eomcl}
\end{align}  
where we introduced the derivative operator $\bar{\cal L}_k$ as
\begin{align}
 & \bar{\cal L}_k := \partial^2_\rho + 3 (1-\varepsilon_1) \partial_\rho 
 +  \frac{e^{-2\rho}}{{\dot{\rho}}^2} k^2 - 3 (\varepsilon_1 +
 \varepsilon_2/2)\,.
\end{align}
Operating $\{(1+\varepsilon_1)\partial_\rho + \bm{k} \cdot \partial_{\sbm{k}}\}$
on Eq.~(\ref{Eq:eomcl}), which commutes with $\bar{\cal L}_k$ up to 
${\cal O}(\varepsilon)$, we obtain
\begin{align}
 & \bar{\cal L}_k \{(1+\varepsilon_1)\partial_\rho + \bm{k} \cdot
 \partial_{\sbm{k}}\} \Theta_{1I} f^I_{\alpha,k} \cr
 & \quad - \left(  3 \frac{\dd \theta}{\dd \rho}  - \frac{V_{\sigma
 s}}{\dot{\rho}^2} + 2 \frac{\dd \theta}{\dd \rho}
 \partial_\rho  \right) (\partial_\rho + \bm{k} \cdot
 \partial_{\sbm{k}}) \Theta_{2I} f^I_{\alpha,k}  =0\,.\cr \label{Exp:reproc}
\end{align}
The mode function for the entropy field $\chi$ is not restricted from the
gauge-invariance condition. However, to reproduce the gauge-invariance
condition (\ref{Exp:GCc1v2}) from Eq.~(\ref{Exp:reproc}), we need to
employ, at the leading order in the slow-roll approximation, the condition
\begin{align}
 &  (\partial_\rho - X^i \partial_{X^i} + D_k)  \Theta_{2I} \psi^I_{\alpha,
 \sbm{k}} = {\cal O}(\varepsilon)\,, \label{Exp:GCc4}
\end{align} 
which almost determines the mode function for $\chi$ to that for the
Bunch-Davies vacuum. Then, the terms in the second line of
Eq.~(\ref{Exp:reproc}) vanish, reproducing Eq.~(\ref{Exp:GCc1v2}) after
the multiplication of $\bar{\cal L}_k^{-1}$. The last two terms in
(\ref{Exp:GCc1v2}) appear as the homogeneous solutions of $\bar{\cal
L}_k$. Following the same argument as in 
Ref.~\cite{IRgauge}, we can also show that the gauge-invariance
conditions and Eq.~(\ref{Exp:GCc4}) sufficiently ensure
that the mode equations are satisfied for all wavenumbers, if they are
satisfied only a particular wavenumber.

\subsection{Primordial non-Gaussianity}
Now we calculate the bi-spectrum for $\gz$, which again becomes the
genuinely gauge-invariant variable for $1/L_{\rm obs} \ll k \ll e^{\rho}H$,
in requesting the gauge-invariance conditions (\ref{Exp:GCc1})\,-\,(\ref{Exp:GCc3}). Substituting Eqs.~(\ref{Exp:dx}),
(\ref{Rel:zetazetanC}), and (\ref{Exp:zetanC}) into Eq.~(\ref{Def:gz}), 
the gauge-invariant curvature perturbation $\gz$ is given by
\begin{align}
 & \gz \Approx \psi + \psi (1+\varepsilon_1) \partial_\rho \psi - \psi
  X^i \partial_{X^i} \psi 
 + \left( \frac{1}{4}\varepsilon_2
 + \xi_2 \right) \psi^2\,, \nonumber\\
\end{align}
where we again abbreviated the several terms in $\delta x^i$  which can give
the non-vanishing contributions for $k \gg 1/L_{\rm obs}$. This is because we
can show that these contributions independently vanish in the squeezed
limit, repeating the same argument as in Ref.~\cite{IRNG}. Note that
after the use of Eq.~(\ref{Exp:GCc3}), the terms with the entropy field
disappear in the expression of $\gz$. The expression of $\gz$ includes
the time-dependent function $\xi_2$, which is restricted from the
request of the consistent quantization~\cite{IRgauge}. Assuming $\xi_2$ is determined
appropriately, we do not discuss the explicit form of $\xi_2$, because this
is not important for our discussions.

Using Eqs.~(\ref{Exp:psi}) and (\ref{Exp:chi}), the gauge-invariant
curvature perturbation $\gz$ can be expanded as
\begin{align}
 & \gz \Approx \sum_{\alpha=1}^2 \Theta_{1I} \psi^I_\alpha + \sum_{\alpha,
 \beta =1}^2 {\cal M}_{\alpha \beta}\,, \label{Exp:gzetac}
\end{align}
where ${\cal M}_{\alpha \beta}$ is defined as
\begin{align}
 & {\cal M}_{\alpha\beta}\! := \Theta_{1I} \psi^I_\alpha \biggl[
 (1+\varepsilon_1) \partial_\rho  -
  X^i \partial_{X^i}\! 
 + \Bigl(\frac{1}{4} \varepsilon_2 + \xi_2  \Bigr) \biggr] \Theta_{1J} \psi^J_\beta\,.
 \nonumber\\[0.5pt] 
\end{align}

The bispectrum for $\gz$ is again given by the three terms in
Eq.~(\ref{Exp:3point}), where only the first two terms give dominant
contributions in the squeezed limit $k_1 \ll k_2 \simeq k_3$.
Using Eq.~(\ref{Exp:gzetac}), the first term in
Eq.~(\ref{Exp:3point}) is given by
\begin{align}
 & \langle \psi_{\sbm{k}_1} \psi_{\sbm{k}_2} \gz_{\sbm{k}_3,2} \rangle
 \nonumber \\[-1pt] & \quad 
\Approx \Theta_{1I} \Theta_{1J} \hspace{-0.4cm}
 \sum_{\alpha,\beta, \alpha',\beta'=1}^2 \hspace{-0.4cm} \langle \psi^I_{\alpha,\,
 \sbm{k}_1} \psi^J_{\beta,\sbm{k}_2} {\cal M}_{\alpha'\beta', \sbm{k}_3}
 \rangle\,,
 \label{Exp:3point1st}
\end{align}
where ${\cal M}_{\alpha\beta, \sbm{k}}$ is the Fourier mode of ${\cal M}_{\alpha\beta}$. Note that 
$
\langle \psi^I_{\alpha,\,\sbm{k}_1} \psi^J_{\beta,\sbm{k}_2} {\cal
M}_{\alpha'\beta', \sbm{k}_3}\rangle
$
gives the non-vanishing contributions only if $\alpha$ and $\beta$ agree with
either $\alpha'$ or $\beta'$, while $\alpha$ and $\beta$ do not
necessarily coincide with each other. We first consider the case with
$\alpha=\beta$, where the non-vanishing contribution is given by
\begin{align}
 &  \langle \psi^I_{\alpha,\,
 \sbm{k}_1} \psi^J_{\alpha,\sbm{k}_2} {\cal M}_{\alpha\alpha, \sbm{k}_3}
 \rangle \cr
 &  \Approx- v^I_{\alpha,k_1} v^J_{\alpha,k_2} \int
 \frac{\dd^3 \bm{p}}{(2\pi)^{3/2}} \int \frac{\dd^3 \bm{X}}{(2\pi)^3}
 \cr & \quad \times\Bigl[ e^{-i(\sbm{k}_1 + \sbm{k}_3) \cdot \sbm{X}}
 \delta^{(3)} (\bm{k}_2+\bm{p}) \psi^*_{\alpha,k_1} D_p
 \psi^*_{\alpha,p} e^{i \sbm{p} \cdot \sbm{X}} \cr
 & \qquad \quad + e^{- i(\sbm{k}_2+\sbm{k}_3) \cdot \sbm{X}}
 \delta^{(3)} (\bm{k}_1+\bm{p}) \psi^*_{\alpha,k_2} X^i
 \partial_{X^i} \psi^*_{\alpha,p} e^{i \sbm{p} \cdot \sbm{X}}
 \Bigr] \cr
 & \qquad + (2\pi)^{-3/2} \delta^{(3)} (\bm{K})\, 
 v^I_{\alpha,k_1} v^J_{\alpha, k_2} \psi_{\alpha, k_2}^* 
 \partial_\rho \psi_{\alpha,\, k_1}^*\,, \label{Exp:Meq}
\end{align}
where $\psi_{\alpha, k}$ is defined as 
$\psi_{\alpha, k} := \Theta_{1I} v^I_{\alpha,k}$ and $\bm{K}$ is defined
as $\bm{K}:= \bm{k}_1+\bm{k}_2+\bm{k}_3$. 
In deriving this expression, we used the gauge-invariance condition
(\ref{Exp:GCc1}) and $\partial_\rho v^I_\alpha = {\cal O}(\varepsilon)$
for $k \ll e^{\rho }H$. 
Next, for the case with $\alpha \neq \beta$, we obtain
\begin{align}
 & \langle \psi^I_{\alpha,\,
 \sbm{k}_1} \psi^J_{\beta,\sbm{k}_2} \left(  {\cal M}_{\alpha\beta,
 \sbm{k}_3} + {\cal M}_{\beta \alpha,\,\sbm{k}_3}  \right)
 \rangle \cr
  &  \Approx- v^I_{\alpha,k_1} v^J_{\beta,k_2} \int
 \frac{\dd^3 \bm{p}}{(2\pi)^{3/2}} \int \frac{\dd^3 \bm{X}}{(2\pi)^3}
 \cr & \quad \times\Bigl[ e^{-i(\sbm{k}_1 + \sbm{k}_3) \cdot \sbm{X}}
 \delta^{(3)} (\bm{k}_2+\bm{p}) \psi^*_{\alpha,k_1} D_p
 \psi^*_{\beta,p} e^{i \sbm{p} \cdot \sbm{X}} \cr
 & \qquad \quad + e^{- i(\sbm{k}_2+\sbm{k}_3) \cdot \sbm{X}}
 \delta^{(3)} (\bm{k}_1+\bm{p}) \psi^*_{\beta, k_2}X^i \partial_{X^i}
 \psi^*_{\alpha ,p} e^{i \sbm{p} \cdot \sbm{X}}
 \Bigr] \cr
 & \qquad + (2\pi)^{-3/2} \delta^{(3)} (\bm{K})\,
 v^I_{\alpha,k_1} v^J_{\beta, k_2} \psi_{\beta, k_2}^* \partial_\rho
 \psi_{\alpha,k_1}^*\,.
 \label{Exp:Mdif}
\end{align}

The calculation of the second term in Eq.~(\ref{Exp:3point}) proceeds
similarly to give
\begin{align}
 & \langle \psi_{\sbm{k}_1} \gz_{\sbm{k}_2,2}  \psi_{\sbm{k}_3} \rangle
 \nonumber \\[-1pt] & \quad 
\approx \Theta_{1I} \Theta_{1J} \hspace{-0.4cm}
 \sum_{\alpha,\beta, \alpha',\beta'=1}^2 \hspace{-0.4cm} \langle \psi^I_{\alpha,\,
 \sbm{k}_1} {\cal M}_{\alpha'\beta', \sbm{k}_2}  \psi^J_{\beta,\sbm{k}_3}
 \rangle\,,
 \label{Exp:3point2nd}
\end{align}
where the non-vanishing contributions are related to those in the first term as
\begin{align}
  &  \Theta_{1I} \Theta_{1J}  \langle \psi^I_{\alpha,\,
 \sbm{k}_1} {\cal M}_{\alpha\alpha, \sbm{k}_2}  \psi^J_{\alpha,\sbm{k}_3}
 \rangle \cr & \quad  
 -  (2\pi)^{-3/2} \delta^{(3)} (\bm{K})\,
 \left| \psi_{\alpha, k_3} \right|^2
  \psi_{\alpha, k_1} \partial_\rho \psi^*_{\alpha, k_1} \cr
 &  \Approx \Bigl[  \Theta_{1I} \Theta_{1J} \langle \psi^I_{\alpha,\,
 \sbm{k}_1}  \psi^J_{\alpha,\sbm{k}_3}  {\cal M}_{\alpha\alpha, \sbm{k}_2}
 \rangle \cr & \quad \,
 -  (2\pi)^{- 3/2} \delta^{(3)} (\bm{K}) 
 \left|\psi_{\alpha, k_3}  \right|^2 \psi_{\alpha, k_1}
 \partial_\rho \psi_{\alpha, k_1}^* \Bigr]^*\hspace{-5pt},\, \label{Exp:Meqv2}
\end{align}
and for $\alpha  \neq \beta$
\begin{align}
  & \Theta_{1I} \Theta_{1J}  \langle \psi^I_{\alpha,\,
 \sbm{k}_1}\left(  {\cal M}_{\alpha\beta,
 \sbm{k}_2} + {\cal M}_{\beta \alpha,\,\sbm{k}_2}  \right)  \psi^J_{\beta,\sbm{k}_3} 
 \rangle \cr
 & \quad -  (2\pi)^{-3/2} \delta^{(3)} (\bm{K})\,
  \left|\psi_{\beta, k_3} \right|^2  \psi_{\alpha,k_1} \partial_\rho
 \psi_{\alpha,k_1}^* \cr
 &\Approx \Bigl[ \Theta_{1I} \Theta_{1J}  \langle \psi^I_{\alpha,\,
 \sbm{k}_1} \psi^J_{\beta,\sbm{k}_3} \left(  {\cal M}_{\alpha\beta,
 \sbm{k}_2} + {\cal M}_{\beta \alpha,\,\sbm{k}_2}  \right)
 \rangle \cr
 & \quad -  (2\pi)^{-3/2} \delta^{(3)} (\bm{K})
  \left|\psi_{\beta, k_3} \right|^2  \psi_{\alpha,k_1} \partial_\rho
 \psi_{\alpha,k_1}^* \Bigr]^*\hspace{-5pt}.
\end{align}

Now, we consider the squeezed limit $k_1 \ll k_2 \simeq k_3$. After some
manipulation, Eq.~(\ref{Exp:Meq}) multiplied by $\Theta_{1I} \Theta_{1J}$
can be expressed as
\begin{align}
 & \Theta_{1I} \Theta_{1J}  \langle \psi^I_{\alpha,\,
 \sbm{k}_1} \psi^J_{\alpha,\sbm{k}_2} {\cal M}_{\alpha\alpha, \sbm{k}_3}
 \rangle \cr
 &  \Approx- \frac{1}{2} \left| \psi_{\alpha, k_1} \right|^2
 \psi_{\alpha, k_2} \psi_{\alpha, |\sbm{k}_1+\sbm{k}_3|}^* \cr
 & \qquad  \qquad  \times 
(2\pi)^{-3/2}(\bm{k}_1-\bm{k}_2+\bm{k}_3)\!\cdot\!
 \partial_{\sbm{K}}\delta^{(3)}(\bm{K})  
 \cr & \qquad - \left| \psi_{\alpha, k_1} \right|^2 \left| \psi_{\alpha,
 k_2} \right|^2 (2\pi)^{-3/2} \bm{k}_1 \cdot \partial_{\sbm{K}}
 \delta^{(3)}(\bm{K}) \cr
 & \qquad + (2\pi)^{-3/2} \delta^{(3)} (\bm{K})\, 
 \psi_{\alpha,k_1} \left|\psi_{\alpha, k_2} \right|^2 
 \partial_\rho \psi_{\alpha,\, k_1}^*. \label{Exp:Meqsq}
\end{align}
(The detailed calculation can be consulted in Ref.~\cite{IRNG}.) The terms with
$\bm{k}_1$ manifestly vanish in the squeezed limit $k_1 \to 0$.
The rest terms on the first and second lines are not suppressed at this moment.
Combining the contribution from the second term of Eq.~(\ref{Exp:3point}),
which satisfies Eq.~(\ref{Exp:Meqv2}), these terms, however, provide 
the factor 
$(\psi_{\alpha, k_2} \psi^*_{\alpha,|\sbm{k}_1+\sbm{k}_3|}-\psi_{\alpha,
|\sbm{k}_1+\sbm{k}_2|}\psi_{\alpha, k_3}^*)$, 
which again vanishes in the squeezed limit, $k_1\to 0$. We therefore
understand that only the term on the last line in Eq.~(\ref{Exp:Meqsq})
yields the non-vanishing contribution in this limit.  Repeating a
similar calculation, we again find that all the terms in Eq.~(\ref{Exp:Mdif})
except for the last one vanish in this limit. Finally, the bispectrum
for $\gz$ in the squeezed limit $k_1 \to 0$ is given by
\begin{align}
 & \langle \gz_{\sbm{k}_1} \gz_{\sbm{k}_2} \gz_{\sbm{k}_3} \rangle \cr
 &\, \approx  2(2\pi)^{-3/2} \delta^{(3)} (\bm{K})\!\! \sum_{\alpha,
 \beta=1}^2\! \left| \psi_{\alpha, k_2} \right|^2  {\rm Re}\! \left[
 \psi_{\beta,k_1} \partial_\rho \psi_{\beta,k_1}^* \right]\,, \cr
 \label{Exp:result}
\end{align}
Note that for $\dd \theta/\dd \rho =0$ the right-hand side of
Eq.~(\ref{Exp:result}) vanishes, reproducing the result in Eq.~(\ref{Exp:3pointR}),
because for this particular case $\psi_{\alpha, k}$ becomes
constant in time at superhorizon scales. By contrast, for 
$\dd \theta/\dd \rho \neq 0$ the adiabatic field does not necessarily
become constant. The result in Eq.~(\ref{Exp:result}) indicates that the
time variation in the local rotation angle can generate the observable
fluctuation, which are not eliminated by gauge transformations.

\section{Conclusion}  \label{Sec:Conclusion}
In this paper we studied the implications of the genuine gauge
invariance in two-field models of inflation. We showed that, likewise in
single field models, if initial quantum states satisfy the
gauge-invariance conditions, the loop diagrams with the adiabatic field
do not yield IR divergences. It is remarkable that the
gauge-invariance conditions, imposed on the adiabatic field, can be
influenced by the entropy field. This is because the interactions between
the adiabatic and entropy fields can generate the additional possibly
divergent terms, which are absent in single field models, and these
contributions are also influenced by gauge effects. For the derivation
of the gauge-invariance conditions, we distinguished the IR
divergences which are relevant to gauge effects from those which are
irrelevant to gauge effects, considering the diagrams up to one-loop order. This
discrimination would become rather complicated if we extend our argument
to higher order in loops. The loop correction of the entropy
field $\langle \delta s\, \delta s \rangle$ then comes to be no longer gauge invariant
due to the contamination of the adiabatic field.

In requesting the gauge invariance in the local universe, we reexamined
the bispectrum for the primordial curvature perturbation. The
conventionally used curvature perturbation $\zeta$ preserves the invariance under
normalizable gauge transformations, but it does not under
non-normalizable gauge transformations. This indicates that the
curvature perturbation $\zeta$ does not preserve the gauge-invariance in
the local universe while this should be preserved in observable
fluctuations. We therefore calculated the tree-level bi-spectrum for
the genuine gauge-invariant curvature $\gz$ to discuss observable fluctuations. In contrast to the result in
single field models, where the genuine gauge-invariant bi-spectrum
completely vanishes in the squeezed limit, in multi-field models, we
still have the non-vanishing contributions in this limit. This is
generated from the time variation in the curvature perturbation at
super-horizon scales, which occurs only if the background trajectory is
curved yielding $\dd \theta/\dd \rho  \neq 0$. This effect can be understood as
follows. While the constant part of the curvature perturbation can be
removed by a local dilatation, which is the non-normalizable gauge
transformation, the time dependent part cannot be removed by gauge
transformations and produce the physical effect. We should emphasize
that our arguments can be applied to the fluctuations in the observable
scales of current measurements. Since the gauge-invariance conditions
derived here are the necessity conditions, precisely speaking, the
genuine gauge invariance of the bi-spectrum for $\gz$ has not
sufficiently proven. It is, however, intuitively reasonable to expect
the non-vanishing physical effects in the presence of the entropy
field. In this paper, we employed the slow-roll approximation. It would be interesting to discuss the
observable effects in models which break the slow-roll approximation,
where large non-Gaussianities are predicted in the conventional
perturbation theory.

At the end of this paper, we add several comments on the IR divergence
from the entropy field. The loop corrections of the entropy field can be
divergent if the entropy field has the scale-invariant or red-tilted
spectrum. This divergence cannot be eliminated also in genuine
gauge-invariant quantities. The IR divergence from the entropy field
should be regularized by a different way from the one for the adiabatic
field. In Ref.~\cite{IRmulti}, we proposed one way to regularize the IR
divergence from the entropy fields. (Some other ways of regularization
were discussed in Refs.~\cite{Bartolo:2007ti,
Riotto:2008mv,Enqvist:2008kt, Burgess:2009bs, Burgess:2010dd}.) We
showed that, if we take into 
account the effects of the quantum decoherence which pick up a unique
history of the universe from various possibilities contained in initial
quantum state, the IR loop corrections of the entropy field no longer
diverge. Therefore, if we consider the decoherence effect, it would be
possible to show the IR regularity of the genuine gauge-invariant
quantities.

\acknowledgments
Y.~U. would like to thank the hospitality of the Perimeter
Institute during the workshop ``IR Issues and Loops in de Sitter Space.''
Y.~U. would also like to thank Robert Brandenberger and Arthur Hebecker for the
hospitalities of McGill university and Heidelberg university. 
The author also acknowledges Jaume Garriga and Takahiro Tanaka for the helpful discussions. 
Y.~U. is supported by
the JSPS under Contact No.\ 21244033, MEC FPA under Contact No.\ 2007-66665-C02, and MICINN
project FPA under Contact No.\ 2009-20807-C02-02.

\end{document}